\documentclass[12pt]{article}

\usepackage{graphicx}
\usepackage{amsmath}
\usepackage{amssymb}
\allowdisplaybreaks

\begin{document}

\baselineskip=17.5pt plus 0.2pt minus 0.1pt

\renewcommand{\theequation}{\arabic{equation}}
\renewcommand{\thefootnote}{\fnsymbol{footnote}}
\makeatletter
\def\CR{\nonumber \\}
\def\pt{\partial}
\def\be{\begin{equation}}
\def\ee{\end{equation}}
\def\bea{\begin{eqnarray}}
\def\eea{\end{eqnarray}}
\def\eq#1{(\ref{#1})}
\def\la{\langle}
\def\ra{\rangle}
\def\hyp{\hbox{-}}


\begin{titlepage}
\title{
\hfill\parbox{4cm}{ \normalsize YITP-10-13}\\
\vspace{1cm} Hopf algebra description of quantum circuits}
\author{
Naoki {\sc Sasakura}\thanks{\tt sasakura@yukawa.kyoto-u.ac.jp}
\\[15pt]
{\it Yukawa Institute for Theoretical Physics, Kyoto University,}\\
{\it Kyoto 606-8502, Japan}}
\date{}
\maketitle
\thispagestyle{empty}
\begin{abstract}
\normalsize
The controlled-NOT gate of qubit quantum circuits is shown to be described in terms of a Hopf algebra.
Accordingly, any qubit quantum circuit can be expressed as the Hopf algebraic computations and unitary
transformations on one qubit.
\end{abstract}
\end{titlepage}

Hopf algebras or quantum groups  \cite{Majid:1996kd} are known to have various applications in physics, 
such as integrable systems, lattice theories, noncommutative spacetimes and quantum gravity.
In view of these examples, Hopf algebras seem to be significant in the physics where group structures in systems 
are deformed by quantum effects or others.
In this letter, as another new example,  I will show that qubit quantum circuits can be described 
by a Hopf algebra\footnote{As a work in a similar direction, see \cite{coecke}.}, 
which can be contrasted with Boolean algebra describing classical logic circuits.  

Let me start with the algebra associated with a classical XOR gate.
The two states of a qubit is described by $|0\rangle$ and $|1\rangle$.
Then the input-output relations of an XOR gate is given by
\begin{equation}
\label{eq:XOR}
\begin{aligned}
|0\rangle\otimes |0\rangle &\rightarrow |0\rangle,\\ 
|0\rangle \otimes|1\rangle &\rightarrow |1\rangle,\\
|1\rangle \otimes |0\rangle &\rightarrow |1\rangle,\\ 
|1\rangle \otimes |1\rangle &\rightarrow |0\rangle,
\end{aligned}
\end{equation}
where $\otimes$ denotes a tensor product.
These relations may be described by a three-index tensor $C_{ab}{}^{c}\ (a,b,c=0 \hbox{ or } 1)$ with values, 
\begin{equation}
\label{eq:valuec}
C_{00}{}^0=C_{01}{}^1=C_{10}{}^{1}=C_{11}{}^{0}=1,\ \hbox{others}=0,
\end{equation}
where the lower and upper indices represent the input and output states, respectively.
Now let me define an algebra with elements $f_a\ (a=0,1)$, the products of which are 
defined by\footnote{Throughout this paper, a pair of upper and lower indices are assumed to be summed over.} 
$f_a \cdot f_b =C_{ab}{}^c f_c$.
From \eq{eq:valuec}, one obtains a commutative associative algebra with products, 
\begin{equation}
\label{eq:algebra}
\begin{aligned}
f_0\cdot f_0&=f_0, \\ 
f_0\cdot f_1&=f_1 \cdot f_0=f_1, \\ 
f_1\cdot f_1&=f_0.
\end{aligned}
\end{equation}
With this algebraic notation, the operation of an XOR gate \eq{eq:XOR} can be considered to be the 
multiplication $m(f_a\otimes f_b)=f_a\cdot f_b$.
As can be seen in \eq{eq:algebra}, the unit is given by
\begin{equation}
\label{eq:unit}
u(1)=f_0.
\end{equation}
The products \eq{eq:algebra} are the same as those of the group $Z_2$.
For the case of quantum circuits, however, this structure must be treated as an algebra but not a group, 
since linear superposition of states, i.e. $c^a f_a\ (c_a\in {\bold C})$, must properly be treated in quantum circuits.

\begin{figure}
\center{\includegraphics[scale=.7]{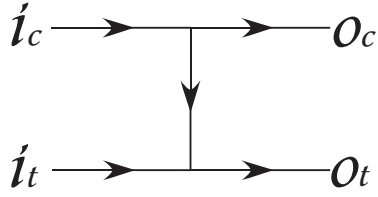}}
\caption{A CNOT gate can be represented by a combination of two three-vertices, which represent the tensor $C$ (the lower vertex) 
and $\tilde C$ (the upper vertex), respectively. The inward and outward oriented lines of each vertex represent
the lower and upper indices of each tensor.}
\label{fig1}
\end{figure}

Let me next discuss a Controlled-NOT (CNOT) gate of qubit quantum circuits. 
The input-output relations are given by 
\begin{equation}
\label{eq:inoutCNOT}
\begin{aligned}
|0\rangle \otimes |0\rangle &\rightarrow& |0\rangle \otimes |0\rangle,\\ 
|0\rangle \otimes |1\rangle &\rightarrow& |0\rangle \otimes |1\rangle,\\ 
|1\rangle \otimes |0\rangle &\rightarrow& |1\rangle \otimes |1\rangle,\\ 
|1\rangle \otimes |1\rangle &\rightarrow& |1\rangle \otimes |0\rangle,
\end{aligned}
\end{equation}
where the former and latter states in the both sides of the arrows denote the control and target states, respectively.
To describe these relations in an algebraic manner, let me introduce another three-index tensor 
$\tilde C_{a}{}^{bc}$ with values,
\begin{equation}
\label{eq:ctvalue}
\tilde C_{0}{}^{00}=\tilde C_{1}{}^{11}=1, \ \hbox{others}=0.
\end{equation}
Then the four-index tensor defined by
\begin{equation}
\label{eq:k}
H_{i_c\, i_t}{}^{o_c\, o_t}=\tilde C_{i_c}{}^{o_c\, a} C_{a\, i_t}{}^{o_t}
\end{equation} 
properly represents the process 
$|i_c\rangle \otimes |i_t\rangle \rightarrow |o_c\rangle \otimes |o_t\rangle$ in \eq{eq:inoutCNOT}.
The tensor $H_{i_c\, i_t}{}^{o_c\, o_t}$ can graphically be described as in Fig.\ref{fig1}, where
the tensors $C$ and $\tilde C$ are represented by three-vertices. 
In fact, this figure is very similar to the graphical representation of a CNOT gate 
often appearing in the literatures of quantum circuits.

The tensor $\tilde C_{a}{}^{bc}$ may define a coalgebra by $\Delta(f_a)=\tilde C_{a}{}^{bc}f_b \otimes f_c$.
For \eq{eq:ctvalue}, 
\begin{equation}
\label{eq:coalgebra}
\begin{aligned}
\Delta(f_0)&=f_0\otimes f_0,\\
\Delta(f_1)&=f_1\otimes f_1,
\end{aligned}
\end{equation}
which is a cocommutative coassociative coalgebra.
This $\Delta$ is the same as the standard operation which generates the tensor product of the group elements of $Z_2$. 
However, to take into account the linear superposition of states, \eq{eq:coalgebra} must be treated as a coalgebra, but not
a tensor product of a group.

Hopf algebra is a bialgebra of an algebra and a coalgebra with unit elements, having a kind of inverse called antipode.
It is well known that a finite group algebra with a coalgebra of standard tensor products 
form a cocommutative Hopf algebra\footnote{See Example 1.5.3 of \cite{Majid:1996kd}.}. In the present case,
the counit and the antipode can be defined by
\begin{equation}
\begin{aligned}
\label{eq:counitantipode}
\varepsilon(f_0)&=\varepsilon(f_1)=1, \\
S(f_0)&=f_0,\\
S(f_1)&=f_1,
\end{aligned}
\end{equation}
respectively.
It is straightforward to show that the definitions \eq{eq:algebra}, \eq{eq:unit}, \eq{eq:coalgebra} and \eq{eq:counitantipode}
define a commutative cocommutative Hopf algebra.

In terms of the Hopf algebra, the CNOT gate (or \eq{eq:k}) has an expression,
\begin{equation}
\label{eq:hopfk}
H_{CNOT}=({\rm id}\otimes m)(\Delta \otimes {\rm id}),
\end{equation} 
where id is the identity operation, $\hbox{id}(f_a)=f_a$.
In fact, one can easily check that $H_{CNOT}(f_a\otimes f_b)=H_{ab}{}^{cd} f_c\otimes f_d$.

\begin{figure}
\center{
\includegraphics[scale=.7]{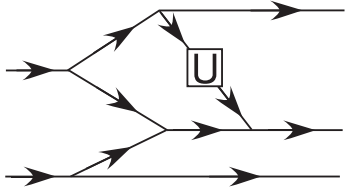}
}
\caption{An example of generalized quantum circuits.}
\label{fig2}
\end{figure}

Any unitary transformation of qubit quantum circuits is known to be realized by a combination of CNOT gates and 
unitary transformations on one qubit \cite{deutsch,Barenco:1995na}. 
Therefore any quantum circuit can be described in terms of the Hopf 
algebra\footnote{The transposition map $\tau(f_a\otimes f_b)=f_b\otimes f_a$ must also be 
introduced to represent crossover of lines.} and unitary transformations on one 
qubit.
On the other hand, however, while quantum computers are defined to be restricted by unitarity, 
the Hopf algebraic computations do not necessarily follow the combination of the CNOT gate \eq{eq:hopfk}, but 
have much more freedom. 
For example, one can consider a circuit in Fig.\ref{fig2}, which is Hopf algebraically represented as
\begin{equation}
(\hbox{id}\otimes m\otimes\hbox{id})(\hbox{id}\otimes U\otimes\hbox{id}\otimes \hbox{id})(\Delta\otimes m\otimes \hbox{id})(\Delta\otimes
\Delta),
\end{equation}
where $U$ is a unitary transformation on one qubit.
This kind of circuits with fewer input states will give only probabilistic outputs for given inputs, and 
may be useful in giving probabilistic predictions based on little information. 
Such ``generalized" quantum computers may be constructed by the same technology which may realize quantum computers
in future.



\end{document}